# Superconducting and Critical Current Properties of NiBi$_3$ Thin Films on Carbon Microfibers and Sapphire

N. Haldolaarachchige, Y. M. Xiong, P. W. Adams and D.P. Young[*]

Department of Physics and Astronomy, Louisiana State University, Baton Rouge, Louisiana 70803, USA

[*]Author to whom correspondence should be addressed:

Email: dyoung@phys.lsu.edu

The superconducting and critical current properties of thin films of NiBi$_3$ formed on the surface of carbon microfibers and sapphire substrates are reported. The NiBi$_3$ coated carbon microfibers were prepared by reacting 7 $\mu$m diameter Ni-coated (~ 80 nm) carbon fibers with Bi vapor, and thin films on sapphire were formed by exposing electron-beam deposited Ni films (~ 40 - 120 nm) to Bi vapor. The microfibers and films show $T_C^{onset}$ = 4.3 K and 4.4 K, respectively, which were slightly higher than that reported for bulk polycrystalline NiBi$_3$. The critical current density ($J_c$) was measured below the transition temperature and is well described by the Ginzburg-Landau power-law: $\left[1-\left(T/T_c\right)^2\right]^{3/2}$, which gives the extrapolated value of $J_c(0) \approx 5.26 \times 10^5 \, \text{A}/\text{cm}^2$.

## I. INTRODUCTION

The existence of superconductivity in compounds containing ferromagnetic elements is very interesting from both an experimental and theoretical point of view. Typically, magnetic elements incorporated into intermetallic superconductors act quite efficiently to destroy the superconductivity. However, the coexistence of these two phenomena was discovered several decades ago,[1-4] and still today the study of magnetism with superconductivity remains at the forefront of condensed matter physics. Ferromagnetic superconductors provide a class of materials to study the effects of magnetic spin fluctuations, which appear to be a salient feature in establishing the superconducting state. This mechanism has been applied to many different classes of materials, including heavy-fermion superconductors,[5] high-$T_c$ cuprates[6] and the recently discovered FeAs-based superconductors.[7]

Previous research showed $NiBi_3$ is a stable intermetallic superconductor with $T_c$ = 4.05 K,[8, 9] crystallizing in the orthorhombic $CaLiSi_2$-type structure with space group *Pnma*.[10] However, only a few reports have discussed the physical properties of this compound.[9-11] Recently, Pineiro, *et al.*[12] have reported the coexistence of ferromagnetism and superconductivity in $NiBi_3$, which is intriguing, but also in conflict with previous reports suggesting a non-magnetic ground state.[10] Kumar *et al.*[13] have done a detailed analysis, which includes an experimental and theoretical study of possible ferromagnetism in this compound, and concluded that $NiBi_3$ should be non-magnetic. Herrmannsdorfer *et al.*[14] used novel chemical-reaction techniques to produce nanostructured $NiBi_3$ and showed the system becomes ferromagnetic when formed in reduced dimensions. They also showed the superconducting state emerges in the ferromagnetic phase and exhibits an upper critical field that exceeds the Clogston limit.[15] Followed by that Xiangde Zhu *et al.*[16] also showed that the ferromagnetic fluctuations exist on the surface of the single crystal of $NiBi_3$ below 150 K by electron spin resonance results.

Motivated by the results mentioned above, we investigated the superconducting state of $NiBi_3$ in reduced dimensions as flat, thin films on sapphire substrates, and as cylindrical shells on the surface of solid carbon microfibers.[17] The latter geometry is ideal for the measurement of the critical current density, because it eliminates the edge effects. We have previously performed similar measurements on microfibers of $MgCNi_3$, $ZnNNi_3$, $Mo_3Sb_7$, and MoN.[15-17]

## II. EXPERIMENTAL TECHNIQUES

Commercially available carbon microfiber (diameter ~ 7 μm) with a very thin Ni coating (~ 80 nm), which is deposited via a proprietary chemical vapor deposition process, was used to make $NiBi_3$ microfibers. Carbon fibers were placed in a quartz tube with high purity bismuth shot. The evacuated tube was heat treated at different temperatures for different time periods. The best superconducting properties were observed in fibers heated at 700 – 800 $^0$C for 5 minutes in a preheated box furnace. Scanning electron micrographs of reacted microfibers showed an obvious change in the Ni coating due to the reaction with bismuth vapor (Fig. 1a). Thin films of $NiBi_3$ of different thicknesses (40 nm to 120 nm) were prepared by heating a Ni thin film on a sapphire substrate with high purity bismuth shot. The precursor Ni films on sapphire were made via e-beam vacuum deposition of an arc-melted Ni button (99.999% Alfa Aesar). The planar thin films are particularly important, since they don't involve carbon, which could possibly affect the physical properties of the microfibers. Electrical resistance at zero field and magnetoresistance were measured by the standard four-probe technique in a Quantum Design Physical Property Measurement System (PPMS); four platinum wires (0.002 inch diameter) were attached using silver paste. The critical current density of the $NiBi_3$ coated carbon fibers was measured with a four-probe geometry using a standard pulse technique. Currents were driven using pulse durations of 1-2 μs with a duty cycle of 1/1000, and the resulting voltages were measured via a boxcar integrator. Care was taken to ensure that a pulse width and a duty cycle were low enough to avoid significant Joule heating at the contacts.[17]

### III. RESULTS AND DISCUSSIONS

Figure 1a shows a scanning election microscopy (SEM) image of the $NiBi_3$ thin layer (~ 80 nm) on a carbon microfiber. After the reaction, the volume of the fiber slightly increases, providing evidence for the incorporation of Bi on the surface. In the Figure, the solid carbon core and cylindrical shell of $NiBi_3$ are clearly distinguishable. We have performed elemental analysis (EDX) on the surface of the $NiBi_3$ thin layer and confirmed that the Ni:Bi ratio is 1:3.

Figure 1b shows the normalized resistivity of a $NiBi_3$ coated carbon microfiber as a function of temperature. A sharp superconducting transition temperature ($T_C^{onset} = 4.3K$) with a very small relative width ($\frac{\Delta T_c}{T_c} = 0.047$) is observed. This transition is slightly higher than the onset temperature of polycrystalline $NiBi_3$.[12] The low temperature resistivity data could not be fitted with standard Fermi Liquid (FL) theory $\left[\rho_T = \rho_0 + AT^2\right]$, and calculated residual resistivity ratio for the fiber of $\rho_{290K}/\rho_{5K} = 4.52$, which shows the poor metallic behavior and high temperature resistivity variation due to disorder.

Field-dependent resistivity of a $NiBi_3$ coated C-microfiber is shown in the Figure 1c. Magnetic field was applied along the fiber axis for all measurements. A sharp superconducting transition at zero field is clearly identified and is shifted toward lower temperature with increasing field. The upper critical field as a function of temperature and reduced temperature is shown in Figures 1d and 1e. The upper critical field varies linearly with the reduced temperature. The extrapolated upper critical field at $T = 0$ for the fiber is $H_{c2}(0) = 0.6$ T, which is slightly higher than that obtained from polycrystalline samples[12] and smaller than recently reported values on the nano-structured NiBi$_3$.[14] This behavior is in contrast with the enhanced upper critical field values observed in other intermetallic superconducting materials in reduced dimensions.[18-20] The value of the reduced upper critical field $\frac{H_{c2}}{T_c}(microfiber) \sim 0.12$ $T/K$ is comparable to that of bulk polycrystalline NiBi$_3$ $\frac{H_{c2}}{T_c}(polycrystalline) \sim 0.09$ $T/K$. The estimated reduced critical field of the fiber is smaller than that of high $T_c$ superconducting materials and the Chandrasekhar-Clogston or Pauli limit.[15] We have calculated the corresponding coherence length of the fiber to be $\xi(0) = 81 \overset{0}{A}$ by using the Ginzburg-Landau formula for an isotropic three-dimensional superconductor; $H_{c2} = \Phi_0 / 2\pi \xi^2(0)$, where $\Phi_0 = 2.0678 \times 10^9 Oe \overset{0}{A}^2$ is the flux quantum.[21, 22] Thus, the film thickness (~ 80 nm) is much larger than the calculated coherence length, which confirms the superconducting film is in the infinite thickness limit.

Figure 2a shows a semi-logarithmic plot of the resistivity versus temperature for NiBi$_3$ thin films of 120 nm and 40 nm thickness. Figure 2b focuses on the low temperature data and shows the superconducting onset temperature (~ 4.4 K) is the same for both films, which is slightly higher than that of the polycrystalline sample and almost same as the $T_c$ of the microfiber. This suggests that if carbon contamination from the surface of the fiber occurs during synthesis, its effect on the superconducting properties of NiBi$_3$ is small. The value of $T_c$ is almost independent of thickness for the NiBi$_3$ films, however, $T_c$ of other intermetallic superconducting films, such as MgCNi$_3$, did vary with thickness.[20] The residual resistivity ratio of the thin film is about $\rho_{290K}/\rho_{5K} = 1.75$, which is consistent with a poor metallic behavior.

Field dependent resistivity measurements are shown in Figure 2c, where the magnetic field was applied along the surface of the film during all measurements. Clearly, the sharp superconducting transition at zero field shifted toward lower temperature with increasing field. The upper critical field as a function of temperature and reduced temperature are shown in Figures 2d and 2e. The extrapolated upper critical field at $T = 0$ K for the thin film is $H_{c2}(0) = 0.5$ T, which is almost same as that of the bulk polycrystalline sample,[12] but smaller than the recently reported values of the nanostructures of this compound.[14] The reduced upper critical field: $\frac{H_{c2}}{T_c}(thin\ film\ 120nm) = 0.09\ T/K$ is exactly the same as that of bulk polycrystalline NiBi$_3$. We have calculated the corresponding coherence length of the thin film to be $\xi(0) = 90\ \overset{0}{A}$, and thus the film thickness (~ 120 nm) is much larger than calculated coherence length.

Critical current measurements on the NiBi$_3$ microfibers are presented in Figure 3. The critical current was measured between 2 - 3.8 K due to limitations of the PPMS base temperature and $T_c$ of this compound. Care was taken to reduce the pulse width and duty cycles to the point were no hysteresis was observed across the critical current threshold. The dashed line in Figure 3 shows the Ginzburg-Landau (GL) critical current behavior for a homogeneous order parameter; $J_c = \frac{H_c(T)}{3\sqrt{6}\pi\lambda(T)} \propto \left[1-\left(T/T_c\right)^2\right]^{3/2}$, where $H_c$ is the thermodynamic critical field, and $\lambda$ is the London penetration depth.[23] Variation of the critical current agrees well with the GL theory. This agreement suggests that there is not any effect of spin fluctuation into a critical current. However, spin fluctuation in MgCNi$_3$ have shown a strong effect on Critical current, which show clear deviation from the GL theory.[24] The zero temperature critical current density was extrapolated to be $J_c = 5.26 \times 10^5\ A/cm^2$, which is lower than that of MgCNi$_3$ and MoN.[19, 24] However, the critical current density data scale very well with the reduced temperature, which suggests that the low value at zero temperature is an intrinsic property of this material. Additionally, we have calculated the critical current density at zero temperature by using the value of the London penetration depth of the bulk polycrystalline sample $\left(J_c = 87.85 \times 10^5 A/cm^2\right)$, which is one order of magnitude larger than that of the extrapolated value.

## IV. CONCLUSIONS

We have successfully synthesized NiBi$_3$ in the form of thin films on the surface of carbon microfiber and on sapphire substrates and measured their superconducting properties. In addition, we have measured the critical current density of the microfibers. The superconducting transition temperature is slightly higher than that of the bulk polycrystalline samples. The critical current behavior is well described by Ginzburg-Landau theory, returning a critical exponent of 3/2. The critical current density at zero temperature is lower than the reported critical current densities of some other intermetallic superconducting compounds also formed on microfibers. However, the current density is comparable with that calculated from the London penetration depth of NiBi$_3$. Finally, we see no direct evidence in the behavior of the critical current density that suggests the existence of ferromagnetic spin fluctuations in NiBi$_3$ at low temperature.

## V. ACKNOWLEDGEMENTS

DPY acknowledges the NSF grant no. DMR-1005764, and PW acknowledges the DOE grant no. DE-FGOZ-07ER46420. NH acknowledges Dr. Amar B. Karki for useful discussions.

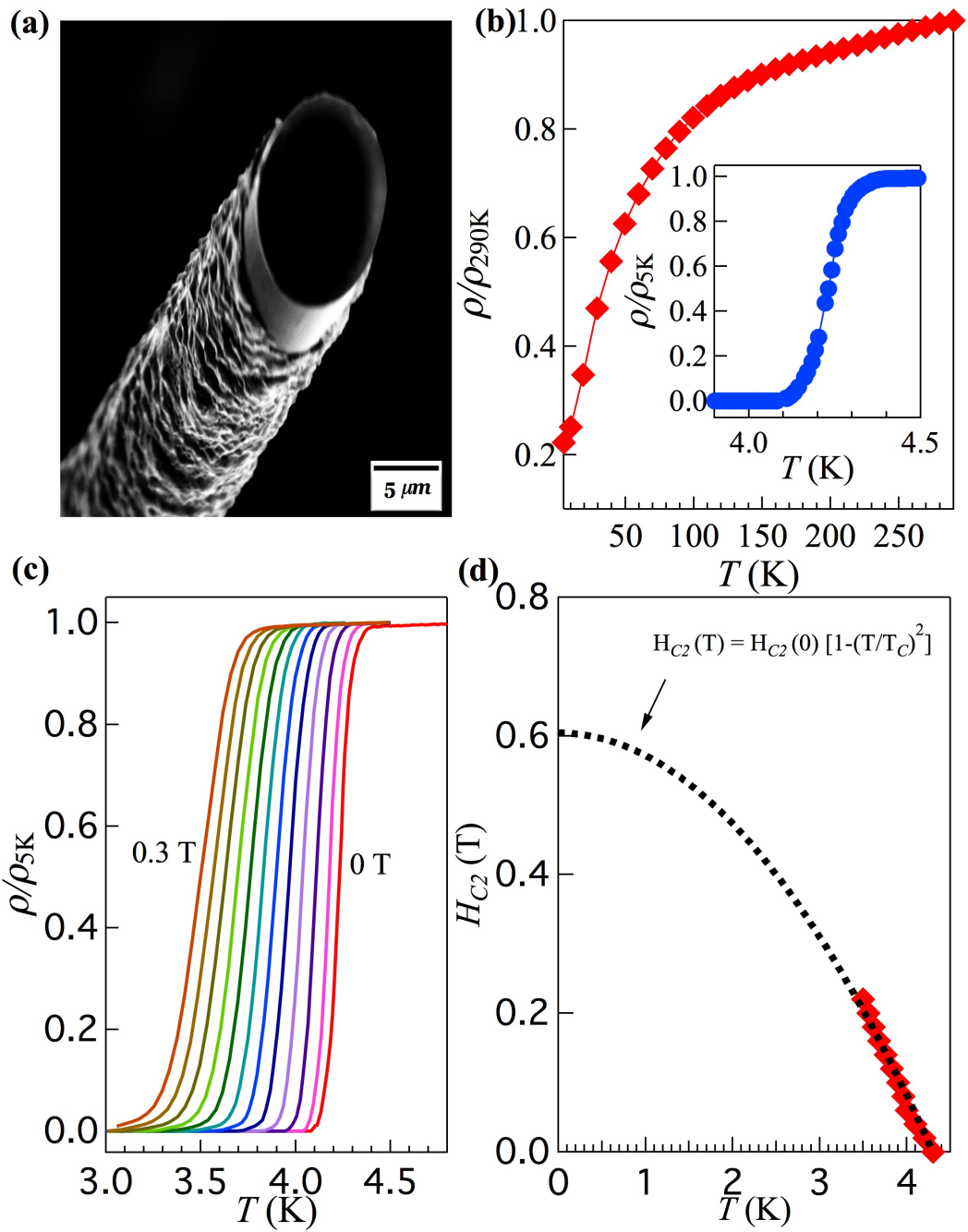

Fig 1 (Color online) (a) SEM image of a single carbon microfiber coated with a thin film of NiBi$_3$, (b) Semi-logarithmic plot of the resistivity versus temperature (inset shows the superconducting transition), (c) Field dependent resistivity measurements, (d) Upper critical field as a function of temperature of a NiBi$_3$ coated carbon microfiber.

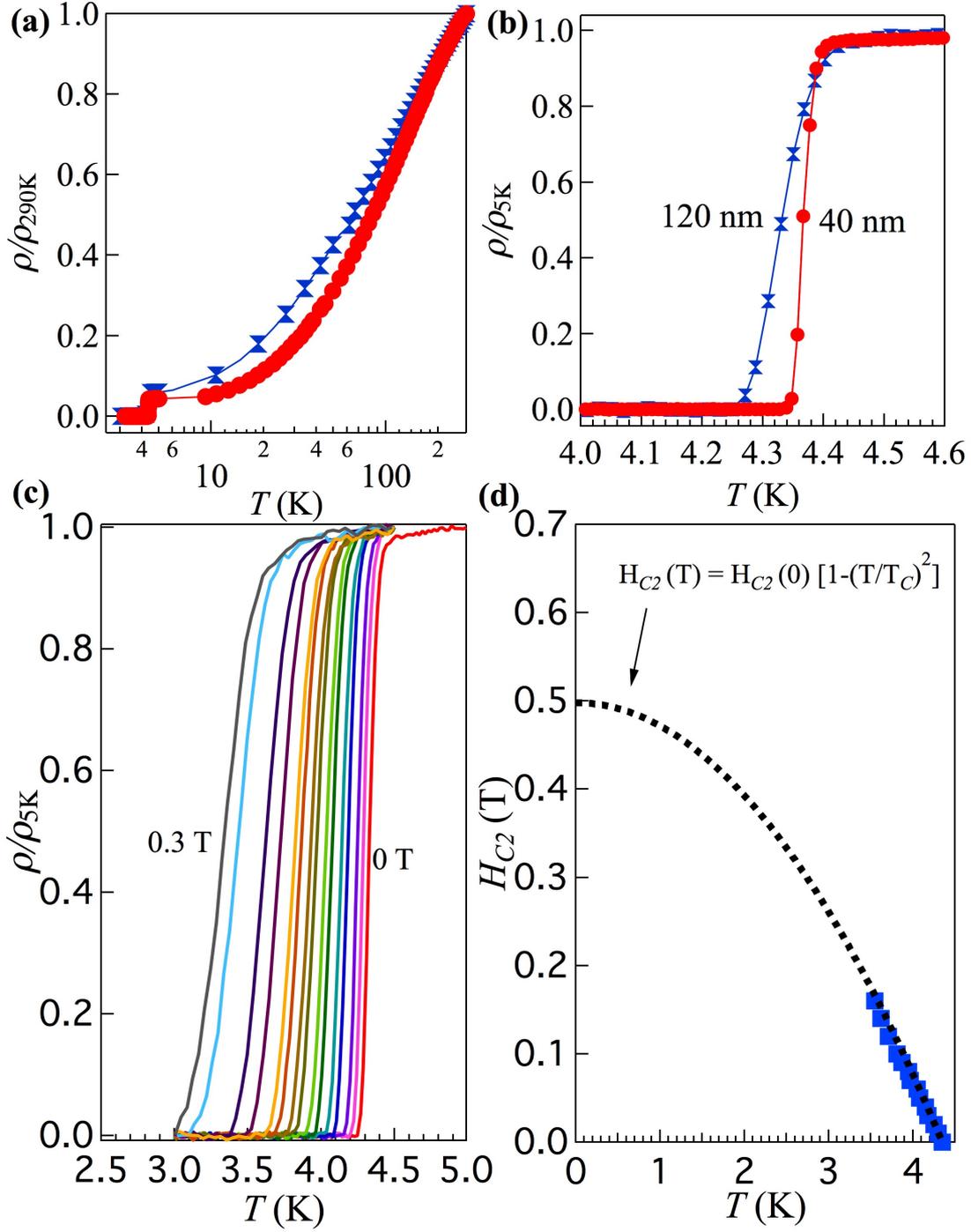

Fig 2 (Color online) (a) Semi-logarithmic plot of resistivity versus temperature for NiBi$_3$ thin films, (b) Superconducting transition temperature, (c) Field dependent resistivity measurements, (d) Upper critical field as a function of temperature of NiBi$_3$ thin film on sapphire substrate.

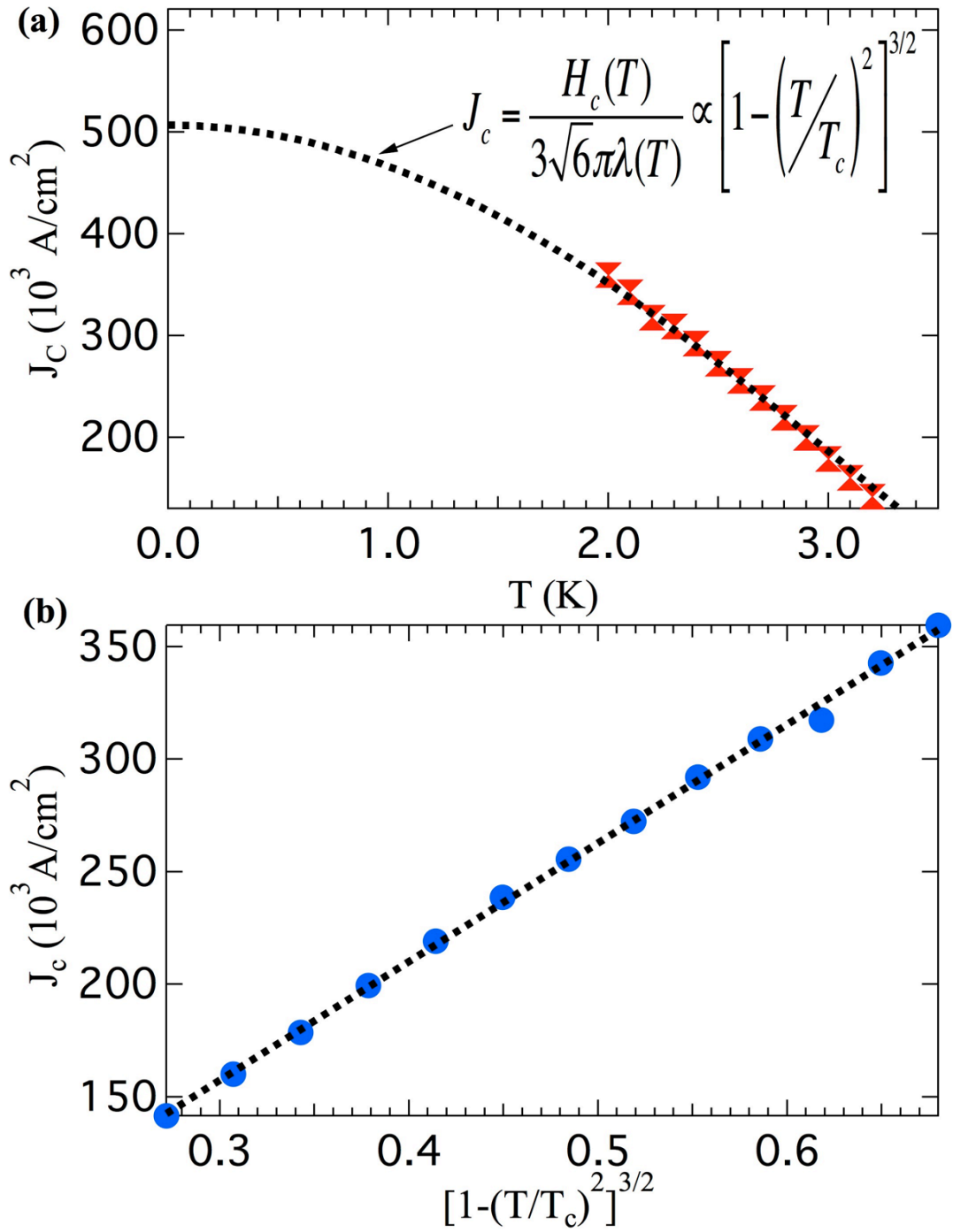

Fig 3 (Color online) Critical current measurement of a single NiBi$_3$ coated carbon microfiber as a function of temperature (a) and reduced temperature (b). The dotted line in (a) is a fit to GL theory as described in the text, and in (b) it is a linear fit.